\documentclass[preprintnumbers,amsmath,amssymb,prd,floatfix,12pt,superscriptaddress,nofootinbib]{revtex4}
\usepackage{graphicx}
\usepackage{epsfig}
\usepackage{bm}
\usepackage{amsfonts}

\begin{document}

\title{Dynamics and Thermodynamics of (2+1)-Dimensional Evolving Lorentzian
Wormhole}
\author{\textbf{M. Umar Farooq}}
\email{mfarooq@camp.nust.edu.pk}
\affiliation{Center for Advanced Mathematics and Physics, National University of Sciences
and Technology, H-12 Campus, Islamabad, Pakistan}
\author{\textbf{M. Akbar}}
\email{makbar@camp.nust.edu.pk}
\affiliation{Center for Advanced Mathematics and Physics, National University of Sciences
and Technology, H-12 Campus, Islamabad, Pakistan}
\author{\textbf{Mubasher Jamil}}
\email{mjamil@camp.nust.edu.pk}
\affiliation{Center for Advanced Mathematics and Physics, National University of Sciences
and Technology, H-12 Campus, Islamabad, Pakistan}

\begin{abstract}
\textbf{Abstract:} In this paper we study the relationship between the
Einstein field equations for the (2+1)-dimensional evolving wormhole and the
first law of thermodynamics. It has been shown that the Einstein field
equations can be rewritten as a similar form of the first law of
thermodynamics at the dynamical trapping horizon (as proposed by Hayward)
for the dynamical spacetime which describes intrinsic thermal properties
associated with the trapping horizon. For a particular choice of the shape
and potential functions we are able to express field equations as a similar
form of first law of thermodynamics $dE=-TdS+WdA$ at the trapping horizons.
Here $E=\rho A$, $T=-\kappa /2\pi $, $S=4\pi \tilde{r}_{A}$, $W=(\rho -p)/2$%
, and $A=\pi \tilde{r}_{A}^{2}$, are the total matter energy, horizon
temperature, wormhole entropy, work density and volume of the evolving
wormhole respectively.\newline
\end{abstract}

\maketitle

\newpage

\section*{Introduction}

Black holes and wormholes solutions to the Einstein field equations have
been studied to investigate the various physical properties of these objects
\cite{BWH}. In the past two decades, enormous work has been done in the
revival of classical wormhole solutions. Historically, the idea of a
wormhole was suggested by Flamm \cite{Flam} by means of the standard
embedding diagram. Later on a similar construction was attempted by Einstein
and Rosen, so-called the Einstein-Rosen bridge \cite{ERB}. Subsequently, it
was shown that the bridge, also called a Schwarzschild wormhole is really a
black hole \cite{Fuller}. But these wormholes were not traversable. Interest
in the wormholes, which were both traversable and stable was stimulated by
the seminal work of Morris and Thorne \cite{Morris1}. A static and
spherically symmetric wormhole possesses intriguing geometrical structure
having a throat that flares out in two opposite directions. The throat
connects two arbitrary regions of the same spacetime or two distinct
spacetimes. There are several reasons that arouse interest in these peculiar
geometries. One of them is the possibility that these spacetimes could be
thought of as time machines \cite{Morris2} which violate Hawking's
chronology protection conjecture. Since the throat has the tendency to get
closed in a very short time (of the order of Planck time), thereby
restricting the time travel possibility. Another reason supports the
possibility of motion of a particle through the throat. Since topologically,
wormholes spacetimes are the same as that of black holes, but a minimal
surface called the throat of wormhole is maintained in the time evolution,
so that a traveler could pass through it in both directions. This could
mainly be possible in the presence of an exotic matter (a matter which
violates the null energy condition (NEC))\ which holds up the wormhole
structure and keeps the wormhole throat open.

In the light of above discussion, it is inevitable to seek a realistic
matter which supports these exotic spacetimes. Recently, the astrophysical
observations suggested that the cosmological fluid violates the NEC \cite%
{Riess}. In the context of cosmology, the phantom energy with the equation
of state $\omega <-1$ is the most obvious choice that violates the NEC and
is expected to be the source that could sustain the wormholes \cite{SVS}.
However, it has been argued that a wormhole could be constructed by letting
the accretion of phantom energy onto black holes \cite{Hay}. On the other
hand, a wormhole may be converted to a black hole as the exotic matter gets
evaporated from the wormhole's throat. So with this discussion, we come to
the point that the wormholes are not merely the mathematical toy spacetime
models with interest only for science fiction movies, but also play as a
model of plausible physical realities that might exist in the very spacetime
fabric of our real universe. The understanding of this reality induced a
revival in the study of wormhole spacetimes, particularly the consideration
of accretion of phantom energy can significantly widen the radius of the
throat \cite{Jamil1}. Further, it has been argued that a wormhole could lead
to an inflationary universe by absorbing large amount of exotic matter \cite%
{Romen}.

Most of the efforts regarding wormhole spacetimes have been devoted to study
static configurations that must satisfy some specific properties in order to
be traversable. However, one can study the wormhole configurations that are
time dependent, such as rotating wormholes \cite{Teo} or wormholes in a
cosmological set up have been discussed in detail in \cite{Kar,Lobo,AVB}.

In this paper, our prime aim is to discuss the thermodynamical feature of
Einstein's field equations for the (2+1)-dimensional evolving wormhole
spacetime. Since in view of global properties, wormholes are quite distinct
from the black holes. But according to Hayward, if one considers the local
properties, both (black holes and wormholes) can be characterized \ by the
presence of marginal (marginally trapped) surfaces, and certainly may be
defined in terms of trapping horizons, which are nothing but hypersurfaces
foliated by marginal surfaces \cite{SA1,SA2,SA3}. So under this
consideration (by defining trapping horizon), he constructed a formalism to
explain the thermodynamical properties of spherical, dynamical black holes
\cite{SA4}. In the case of wormholes, the idea to study the thermodynamics
can be motivated by the fact that even though the definition of an event
horizon is no longer possible for the evolving wormhole, but we can still
introduce the concept of a trapping horizon for these objects. Hence
introducing the trapping horizon for wormhole spacetimes allows one to study
these peculiar geometries unifying them with black holes \cite{Hay}.
Therefore, the idea that wormholes may show some characteristics and
properties which are parallel to those already found in black holes, seems
to be quite natural, including in particular "wormhole thermodynamics" \cite%
{Hay}. Recently, the authors \cite{Prado} have shown that one can construct
three laws of thermodynamics for Lorentzian wormholes by using trapping
horizons. They demonstrated that these laws are related with a thermal
phantom-like radiation process coming from the wormhole throat.

In literature, the work on wormhole spacetimes in lower \cite{Perry} and
higher dimensions have been studied by many authors. For instance, the
Euclidean wormholes have been considered by Gonzales-Diaz and by Jianjun and
Sicong \cite{diaz}. The discussion on Lorentzian wormholes in the
n-dimensional Einstein gravity or Einstein-Gauss-Bonnet theory of
gravitation is available in \cite{Gerard} while the author \cite{rami}
discussed the interesting features of static wormhole in higher dimensional
cosmology by considering the scale factor $a(t)$. Mostly the authors
construct the wormholes solutions in (3+1)-dimensional gravity. However,
there are few solutions of wormhole in (2+1) dimensional gravity which is a
covariant theory of gravity that has a great simplicity when compared with
general relativity. This theory has been applied to study some quantum
aspects of gravity \cite{Carlip}. As in the case of (2+1)-dimensional
Banados, Teitelboim and Zanelli (BTZ) black hole \cite{btz}, several authors
have shown interest in wormholes in (2+1) dimensional gravity. Delgaty et
al. \cite{Del} studied the characteristics of traversable wormholes in the
(2+1) dimensional gravity with cosmological constant. Aminneborg et al. \cite%
{amin} compared the properties of wormhole with the characteristics of black
hole while Kim et al. \cite{Kim1} gave two specific solutions of wormhole
taking (2+1) dimensional gravity with a dilatonic field. Further Rahaman et
al. \cite{rahman} and Jamil and Farooq \cite{jami} constructed the phantom
wormholes in the lower dimensional (2+1) gravity. In a similar fashion the
thermodynamics of lower dimensional gravity has been studied to generalize
the deep relation between gravity theories and thermodynamics \cite{btz1}.
To study this relation in a wide range of spacetime geometries, it is
inevitable to extend it to the (2+1)-dimensional wormhole.

This paper is structured as follows: The next section contains some basic
ideas about the Morris and Thorne wormhole and gives the definition of the
evolving wormholes. Then we discuss the spatial geometry of the
(2+1)-dimensional evolving wormhole. In Section three we discuss the
dynamics of the evolving wormhole at the trapping horizon by introducing the
phantom energy as a perfect fluid. Section four deals with the thermal
interpretation of the field equations of (2+1)-dimensional evolving wormhole
at the trapping horizon. Finally we present the conclusion in the last
section.

\section*{Preliminaries}

\subsection*{The Morris-Thorne Wormhole}

Before going over the discussion of evolving wormhole, let us first recall
some properties of static case. Morris and Thorne \cite{Morris1,Morris2}
work out a general, static and spherically symmetric traversable wormhole
described by the metric%
\begin{equation}
ds^{2}=-e^{2\Phi (r)}dt^{2}+\Big[\frac{dr^{2}}{1-\frac{b(r)}{r}}%
+r^{2}(d\theta ^{2}+\sin ^{2}\theta d\phi ^{2})\Big],
\end{equation}%
where $\Phi (r)$ denotes the redshift function and $b(r)$ is the shape
function. The redshift function $\Phi (r)$ must be finite throughout the
spacetime in order to ensure the absence of horizons and singularities. The
minimum radius $r_{0}$ corresponds to the throat of the wormhole, where $%
b(r_{0})=r_{0}$ and the embedded surface is verticale. The proper radial
distance is defined by%
\begin{equation}
l(r)=\pm \int_{r_{0}}^{r}\frac{dr}{\sqrt{1-b(r)/r}},
\end{equation}%
and it must be finite throughout the wormhole spacetime, i. e. $b(r)/r\leq 1$
for $r\geq r_{0}.$ Here $\pm $ signs represent the asymptotically flat
regions which are connected by the wormhole's throat. The equality sign in $%
b(r)/r\leq 1$ holds only at the throat. The both functions ($\Phi (r)$ and $%
b(r)$) must tend to a constant value as the radial coordinate goes to
infinity. The shape function must satisfy the asymptotic flatness condition,
i. e. as $l\rightarrow \pm \infty $ (or equivalently, $r\rightarrow \infty $%
) then $b(r)/r\rightarrow 0.$ With these constraints we are able to see
through an analysis of the embedding diagram of (1) in a Euclidean space,
two asymptotically flat sections connected by a throat.

\subsection*{Evolving Lorentzian Wormhole}

In the cosmological set up, the evolving wormhole spacetime may be obtained
by a simple generalization of the Morris and Thorne metric (1) to a time
dependent metric given by \cite{Romen,Kar,AVB,Catal1,Catal2,AYK}

\begin{equation}
ds^{2}=-e^{2\Phi (t,r)}dt^{2}+a^{2}(t)\Big[\frac{dr^{2}}{1-\frac{b(r)}{r}}%
+r^{2}(d\theta ^{2}+\sin ^{2}\theta d\phi ^{2})\Big],
\end{equation}%
where $a(t)$ is the scale factor of the universe. Worth noticing point is
that the essential characteristics required for a wormhole geometry are
still lies in the spacelike section. Notice that if $b(r)\rightarrow 0$ and $%
\Phi (t,r)\rightarrow 0$ the wormhole metric (3) becomes the flat
Friedmann-Robertson-Walker (FRW) metric, and as $a(t)\rightarrow $\textit{%
constant }it reduces to the static wormhole metric (1). For the construction
of an evolving wormhole, one has to choose some ansatz for the redshift
function $\Phi (t,r),$ the shape function $b(r)$ or the scale factor $a(t)$
and the others are determined by employing some physical conditions$.$ For
instance the author \cite{Romen} considered an exponential scale factor in
order to explore the possibility that inflation might provide a natural
mechanism for the enlargement of an initially small (possibily
submicroscopic) wormhole to macroscopic size. Since we will be dealing with
the dynamical spacetime in lower dimension, so by using the ansatz $\Phi
(t,r)=0$ and $b(r)=r_{0}^{2}/r,$ (where $r_{0}$ is a finite radius of the
wormhole's throat) as given by \cite{Morris1}, the metric for
(2+1)-dimensional evolving wormhole is given as%
\begin{equation}
ds^{2}=-dt^{2}+a^{2}(t)\Big[\frac{dr^{2}}{1-\frac{r_{0}^{2}}{r^{2}}}%
+r^{2}d\theta ^{2}\Big].
\end{equation}

\subsection*{Spatial Geometry of the (2+1)-Dimensional Evolving Wormhole}

To investigate the geometric nature of the (2+1)-dimensional wormhole given
by the metric (4), we embed the spatial geometry of (2+1) wormhole into a
flat 3-dimensional Euclidean space $R^{3}$ \cite{Morris1} together with the
study of expansion factor $a(t)$ involved in our case$.$ So using the
spherical symmetry we can set an equatorial slice ($\theta =\pi /2$) at some
fix instant of constant time $t=t_{0}$ which implies $dt=0.$ Hence the
metric on the resulting two-surface is%
\begin{equation}
ds^{2}=\frac{a_{0}^{2}dr^{2}}{1-r_{0}^{2}/r^{2}}+a_{0}^{2}r^{2}d\phi ^{2},
\end{equation}%
where $a(t_{0})=a_{0}$ is the value of the scale factor at $t=t_{0}$. By
substituting $a_{0}r=\bar{r}$ which implies $a_{0}^{2}dr^{2}=d\bar{r}^{2},$
the above metric (5) can be rewritten as%
\begin{equation}
ds^{2}=\frac{d\bar{r}^{2}}{1-a_{0}^{2}r_{0}^{2}/\bar{r}^{2}}+\bar{r}%
^{2}d\phi ^{2},
\end{equation}%
where we set $a_{0}r_{0}=\bar{r}_{0}$ is the throat radius of the evolving
wormhole at a specific time $t=t_{0}$. The throat radius $\bar{r}%
_{0}\gtreqless r_{0}$ accordingly $a_{0}\gtreqless 1.$ The 3-dimensional
Euclidean space at $t=t_{0}$ can be written in cylindrical coordinates $%
(r,\phi ,z)$ as%
\[
ds^{2}=d\bar{z}^{2}+d\bar{r}^{2}+\bar{r}^{2}d\phi ^{2},
\]%
or%
\begin{equation}
ds^{2}=(1+d\bar{z}^{2}/d\bar{r}^{2})d\bar{r}^{2}+\bar{r}^{2}d\phi ^{2}.
\end{equation}%
In this study we use inflation to enlarge an initially small (possibly
submicroscopic) wormhole. This inflated wormhole will have the same overall
size and shape relative to the initial $\bar{z},\bar{r},\phi $ as the
initial wormhole had relative to the initial $z,r,\phi $ embedding space
coordinate system. This is just because we are considering a series of
embedding spaces, each corresponds to a particular value of $t=$ \textit{%
constant} whose $z,r$ coordinates scale with time. After comparing the
metric (6) and (7), one can readily workout the equation of the embedded
surface as%
\begin{equation}
\frac{dz}{d\bar{r}}=\pm \frac{\bar{r}_{0}}{\sqrt{\bar{r}^{2}-\bar{r}_{0}^{2}}%
},
\end{equation}%
equivalently%
\begin{equation}
\frac{dz}{dr}=\pm \frac{r_{0}}{\sqrt{r^{2}-r_{0}^{2}}}.
\end{equation}%
The Eqs. (6) and (7) imply that the evolving wormhole will remain the same
size in the $\bar{z},\bar{r},\phi $ coordinates. So from equation (9) it is
evident that at the throat radius $r=r_{0},$ the embedded surface is
vertical for which $\frac{dz}{dr}\rightarrow \infty .$ As $r\rightarrow
\infty $, $\frac{dz}{dr}\rightarrow 0$ implies the embedded geometry is
asymptotically flat. The equation (9) represents two embedded surfaces above
and below the $r$-axis and these two surfaces join at $r=r_{0}.$ To
visualize the embedded diagram we integrate equation (9) which yields
\begin{equation}
z(r)=\pm r_{0}\left[ \ln |r+\sqrt{r^{2}-r_{0}^{2}}|-\ln |r_{0}|\right] .
\end{equation}%
In Figures (1-3), we have plotted the surface of revolution about the z-axis
for a fixed radius $r_{0}$ and different choices of the scale factor $%
0<a_{0}<1$, $a_{0}=1$ and $a_{0}>1$, which correspond to respectively
contracting, static and expanding wormholes.

\begin{figure}[!hp]
\scalebox{0.3}{\includegraphics{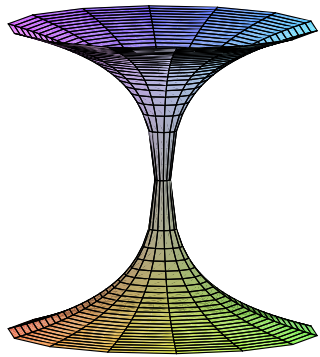}}
\caption{The full visualization of the surface generated by the rotation of
the embedded curve about the vertical axis. Chosen parameters are $r_0=1$
and $r=1...10$}
\end{figure}

\begin{figure}[!hp]
\scalebox{0.4}{\includegraphics{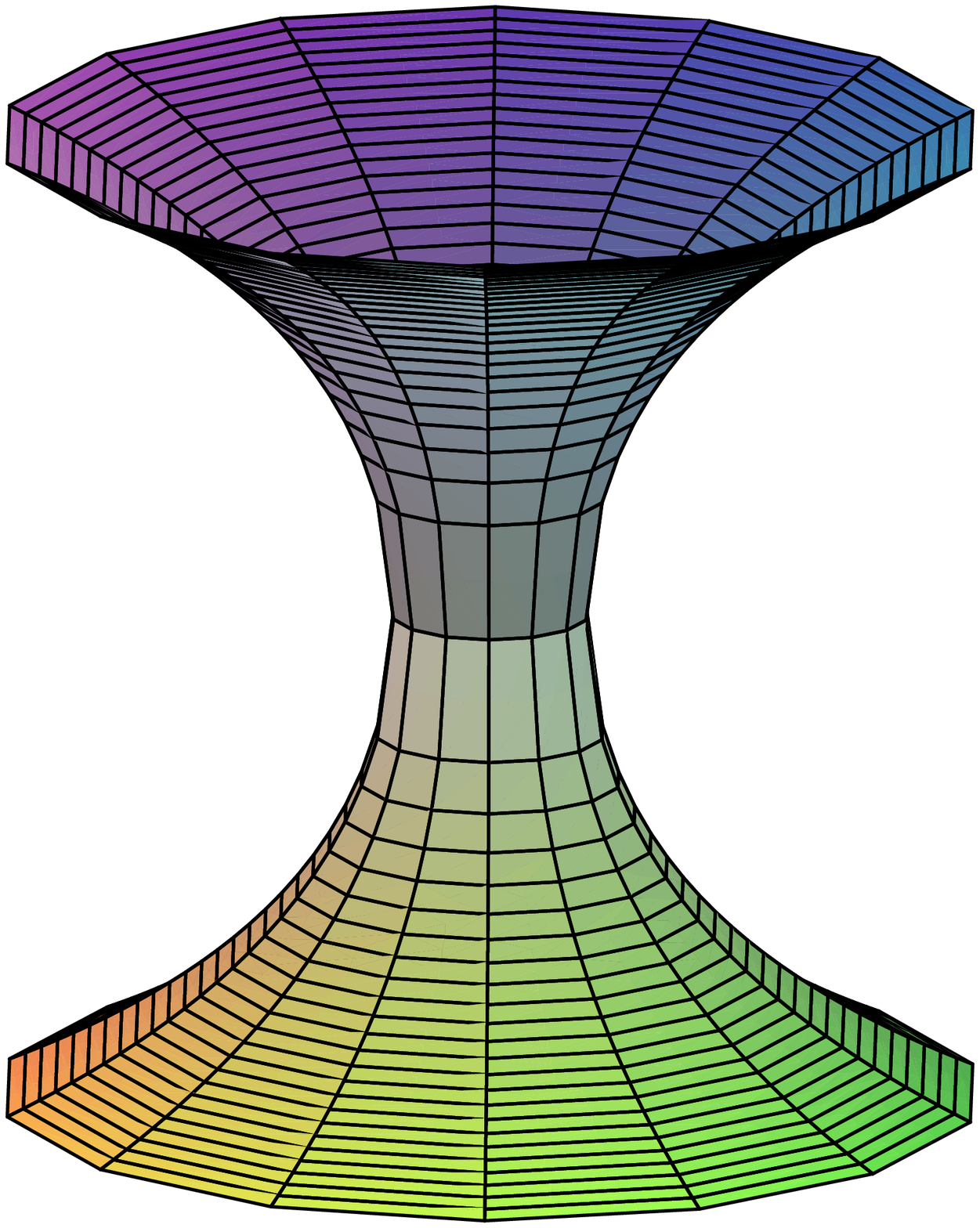}}
\caption{The full visualization of the surface generated by the rotation of
the embedded curve about the vertical axis. Chosen parameters are $r_0=2$
and $r=2...10$}
\end{figure}

\begin{figure}[!hp]
\scalebox{0.5}{\includegraphics{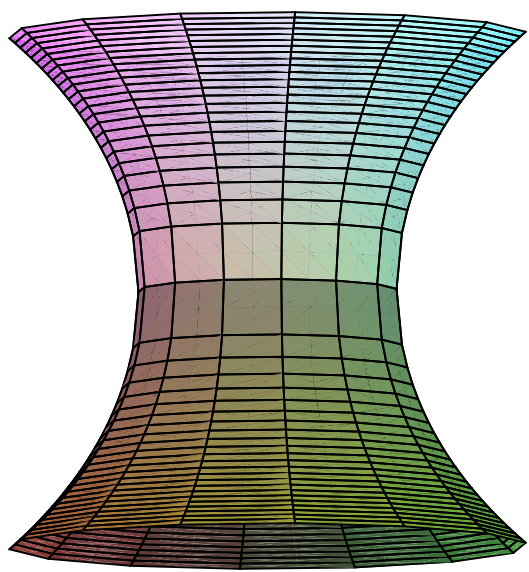}}
\caption{The full visualization of the surface generated by the rotation of
the embedded curve about the vertical axis. Chosen parameters are $r_0=4$
and $r=4...10$}
\end{figure}

One can define the proper radial distance $l(r)=\pm \int\limits_{r_{0}}^{r}%
\frac{a(t)dr}{\sqrt{1-b(r)/r}}$, which must be well behaved everywhere \cite%
{Morris1}. For our model, this parameter $l$ at $t=t_{0}$ for the upper part
of the wormhole $z>0$ is
\begin{equation}
l(r)=a_{0}\sqrt{r^{2}-r_{0}^{2}},
\end{equation}%
and for the lower part $z<0$
\begin{equation}
l(r)=-a_{0}\sqrt{r^{2}-r_{0}^{2}}.
\end{equation}%
In Figure 4, we have plotted the proper radial distance against $r$ which
shows that this function $l(r)$ is a well-behaved quantity everywhere. It is
clear from the diagrams that both, the size of the throat and the radial
proper distance between the wormhole mouths increases with the passage of
time. In our case, as mentioned earlier, far from the wormhole mouth the
space is asymptotically flat. On the other hand, if we take the cosmological
constant into account, then the resulting space could be de-Sitter or
anti-de Sitter far from the mouth. In literature different form of scale
factor are available, for instance, the authors \cite{Catal1} introduced a
linear form of scale factor while in \cite{Romen} different exponential
forms of the scale factor are discussed and their cosmological aspects have
also been studied. So the role of the scale factor becomes so important with
clear indications that as the wormhole inflate, its throat size and proper
length inflate along with the surrounding space.

\begin{figure}[!hp]
\scalebox{0.3}{\includegraphics{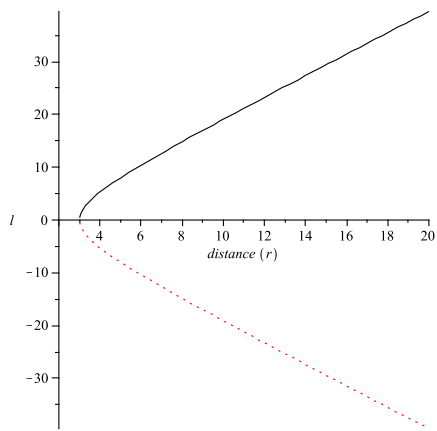}}
\caption{The proper radial distance $l$ is shown. The chosen parameters are $%
a_0=2$ and $r_0=3$.}
\end{figure}

\section*{Dynamics of the Evolving Wormhole at the Trapping Horizon}

This section deals with some dynamical features of the (2+1)-dimensional
evolving wormhole. Since the term event horizon is no longer possible for
the wormhole spacetimes, but still we can introduce for them the term
trapping horizon to discuss the dynamics of these peculiar geometries. It
has been argued that the trapping horizon for a dynamical spacetime is a
causal horizon associated with the gravitational entropy and the surface
gravity \cite{SA4,Bak,Cai}. The concept of trapping horizon for the wormhole
geometries will lead us to explore some physical insights of these weird
objects. So to start with, the spherically symmetric form of the metric (4)
is%
\begin{equation}
ds^{2}=h_{ab}dx^{a}dx^{b}+\tilde{r}^{2}d\theta ^{2},\ \ \ (a,b=0,1)
\end{equation}%
where $x^{0}=t,$ $x^{1}=r,$ and $\tilde{r}=ar$ represents the radius of the
sphere while the two dimensional metric is written as%
\begin{equation}
h_{ab}=\text{diag}\Big(-1,a(t)^{2}\Big(1-\frac{r_{0}^{2}}{r^{2}}\Big)^{-1}%
\Big).
\end{equation}

In this section we revisit some of the terms and notations used in Hayward
formalism based on the dual-null coordinates \cite{Prado,Cai2}. The double
Null coordinates by using metric (4) are constructed and are given by%
\begin{equation}
d\xi ^{+}=-\frac{1}{\sqrt{2}}(dt-\frac{adr}{\sqrt{1-r_{0}^{2}/r^{2}}}),
\end{equation}%
\begin{equation}
d\xi ^{-}=-\frac{1}{\sqrt{2}}(dt+\frac{adr}{\sqrt{1-r_{0}^{2}/r^{2}}}),
\end{equation}%
so with these coordinates, the wormhole metric (13) takes the form%
\begin{equation}
ds^{2}=-2d\xi ^{+}d\xi ^{-}+\tilde{r}^{2}d\theta ^{2}.
\end{equation}%
It is easy to write
\begin{equation}
\partial _{+}=\frac{\partial }{\partial \xi ^{+}}=-\sqrt{2}(\frac{\partial }{%
\partial t}-\frac{\sqrt{1-r_{0}^{2}/r^{2}}}{a}\frac{\partial }{\partial r})
\end{equation}%
\begin{equation}
\partial _{-}=\frac{\partial }{\partial \xi ^{-}}=-\sqrt{2}(\frac{\partial }{%
\partial t}+\frac{\sqrt{1-r_{0}^{2}/r^{2}}}{a}\frac{\partial }{\partial r}),
\end{equation}%
where the minus signs ensure that $\partial _{\pm }$ are future pointing.
One can define the expansions as%
\begin{equation}
\theta _{\pm }=\frac{2}{\tilde{r}}\partial _{\pm }\tilde{r}.
\end{equation}%
Since the sign of $\theta _{+}\theta _{-}$ is invariant, one can say that a
sphere is trapped if $\theta _{+}\theta _{-}>0,$ i. e.%
\begin{equation}
\left( H^{2}\tilde{r}^{4}-\tilde{r}^{2}-a^{2}r_{0}^{2}\right) >0,
\end{equation}%
untrapped if $\theta _{+}\theta _{-}<0$ i. e.%
\begin{equation}
\left( H^{2}\tilde{r}^{4}-\tilde{r}^{2}-a^{2}r_{0}^{2}\right) <0,
\end{equation}%
and marginal if $\theta _{+}\theta _{-}=0$ i. e.%
\begin{equation}
\left( H^{2}\tilde{r}^{4}-\tilde{r}^{2}-a^{2}r_{0}^{2}\right) =0.
\end{equation}%
If the expansion $\theta _{+}>0$ and $\theta _{-}<0$ is locally fixed on an
untrapped sphere, then $\partial _{+}$ and $\partial _{-}$ are also fixed as
the outgoing and ingoing null normal vectors (or the contrary if the
orientation $\theta _{+}<0$ and $\theta _{-}>0$ is considered). A marginal
sphere with $\theta _{+}=0,$ i. e.%
\begin{equation}
-\frac{2\sqrt{2}}{\tilde{r}^{2}}\left( H\tilde{r}-\sqrt{1-a^{2}r_{0}^{2}/%
\tilde{r}^{2}}\right) =0,
\end{equation}%
is future if $\theta _{-}<0,$ i. e.%
\begin{equation}
-\frac{2\sqrt{2}}{\tilde{r}^{2}}\left( H\tilde{r}+\sqrt{1-a^{2}r_{0}^{2}/%
\tilde{r}^{2}}\right) <0,
\end{equation}%
past if $\theta _{-}>0,$ i. e.%
\begin{equation}
-\frac{2\sqrt{2}}{\tilde{r}^{2}}\left( H\tilde{r}+\sqrt{1-a^{2}r_{0}^{2}/%
\tilde{r}^{2}}\right) >0
\end{equation}%
and bifurcating if $\theta _{-}=0,$ i. e.%
\begin{equation}
-\frac{2\sqrt{2}}{\tilde{r}^{2}}\left( H\tilde{r}+\sqrt{1-a^{2}r_{0}^{2}/%
\tilde{r}^{2}}\right) =0.
\end{equation}%
This marginal sphere is outer if $\partial _{-}\theta _{+}<0$
(equivalentally $\partial _{-}\partial _{+}\tilde{r}<0$) i. e.%
\begin{equation}
-\frac{2\sqrt{2}}{\tilde{r}^{2}}\left( \dot{H}+2H-\frac{a^{2}r_{0}^{2}}{%
\tilde{r}^{4}}\right) <0,
\end{equation}%
inner if $\partial _{-}\theta _{+}>0$ (equivalently $\partial _{-}\partial
_{+}\tilde{r}>0$)i. e.%
\begin{equation}
-\frac{2\sqrt{2}}{\tilde{r}^{2}}\left( \dot{H}+2H-\frac{a^{2}r_{0}^{2}}{%
\tilde{r}^{4}}\right) >0
\end{equation}%
and degenerate if $\partial _{-}\theta _{+}=0$ (equivalentally $\partial
_{-}\partial _{+}\tilde{r}=0$) i. e. $\dot{H}+2H-\frac{a^{2}r_{0}^{2}}{%
\tilde{r}^{4}}=0.$ A hypersurface foliated by marginal sphere is called a
trapping horizon and has the same classification as the marginal sphereres.
Hence the expression for the trapping horizon yields%
\begin{equation}
\theta _{+}=\left( H\tilde{r}-\sqrt{1-a^{2}r_{0}^{2}/\tilde{r}^{2}}\right)
=0.
\end{equation}

Let us now consider the Einstein field equations
\begin{equation}
G_{mn}=-\pi T_{mn},\ \ \ (m,n=0,1,2)
\end{equation}%
where $G_{mn}$ is the Einstein tensor and $T_{mn}$ is the energy-momentum
tensor of the matter fields. We have used $G=1/8$ in Eq. (31). One
motivating feature of discussion of the evolving wormhole is the possibility
of sustaining of these geometries by means of exotic matter made out of
phantom energy. The later is used as a main source to explain the late time
accelerated expansion of the universe \cite{Nojiri}. Since this energy
violates the NEC, so it is used as a most appropriate ingredient that could
sustain the wormhole geometries. So we take phantom energy as a perfect
fluid for the evolving wormhole given by%
\begin{equation}
T^{mn}=(\rho +p)u^{m}u^{n}+pg^{mn},
\end{equation}%
where $\rho (t)$ and $p(t)$ are time dependent energy density and pressure
while $u^{m}=(1,0,0)$ is the comoving three velocity of the fluid. The
energy conservation condition $T_{;m}^{mn}=0,$ yields $\dot{\rho}+2H(\rho
+p)=0$. Solving the Einstein field equation (31), in the background of
wormhole geometry (4), one can get by utilizing $G_{0}^{0}$ and $G_{1}^{1}$,
the Friedman-like field equation
\begin{equation}
\dot{H}+\frac{r_{0}^{2}}{a^{2}r^{4}}=-\pi (\rho +p),
\end{equation}%
where $H=\dot{a}/a$, is the Hubble parameter and overdot indicates the
derivative with respect to the cosmic time. The explicit form of the
trapping horizon can be evaluated by using the relation $\theta _{+}=0$,
which after simplification yields
\begin{equation}
H^{2}\tilde{r}_{A}^{4}-\tilde{r}_{A}^{2}+a^{2}r_{0}^{2}=0,\text{ \ \ where }%
\tilde{r}_{A}=a(t)r
\end{equation}%
which is quadratic in $\tilde{r}_{A}^{2}$. It can be seen from Eq. (34) that
when $r_{0}=0$ we have namely a flat FRW universe, and the wormhole trapping
horizon $\tilde{r}_{A}$ takes the same value as the Hubble horizon, $\tilde{r%
}_{A}=1/H$. The Hubble parameter in terms of the wormhole trapping radius is
$H^{2}=(1/\tilde{r}_{A}^{2}-a^{2}r_{0}^{2}/\tilde{r}_{A}^{4})$, and its time
derivative yields
\begin{equation}
\dot{H}=-\frac{\dot{\tilde{r}}_{A}}{H\tilde{r}_{A}^{3}}\Big(1-\frac{%
2a^{2}r_{0}^{2}}{\tilde{r}_{A}^{2}}\Big)-\frac{a^{2}r_{0}^{2}}{\tilde{r}%
_{A}^{4}}.
\end{equation}%
The trapping horizons of the wormhole metric (13) are described by the roots
of the Eq. (34) which yields%
\begin{eqnarray}
\tilde{r}_{A+}^{2} &=&\frac{1+\sqrt{1-4H^{2}a^{2}r_{0}^{2}}}{2H^{2}},
\nonumber \\
\tilde{r}_{A-}^{2} &=&\frac{1-\sqrt{1-4H^{2}a^{2}r_{0}^{2}}}{2H^{2}}.
\end{eqnarray}%
There are three cases depending upon the roots; (a) Two distinct real roots (%
$1-4H^{2}a^{2}r_{0}^{2}>0$) refer as a usual wormhole geometry, (b) two
repeated real roots ($1-4H^{2}a^{2}r_{0}^{2}=0$) called as the `extreme
wormhole' geometry, (c) no real roots ($1-4H^{2}a^{2}r_{0}^{2}<0$) imply the
`naked wormhole'. The notion of nakedness refers to the absence of dynamical
trapping horizon. If we assume that $0<r_{0}^{2}\ll 1$, and neglecting $%
O(r_{0}^{4})$, it is possible to simplify the expressions for $\tilde{r}%
_{A+} $ and $\tilde{r}_{A-}$, which give
\begin{equation}
\tilde{r}_{A+}^{2}=\frac{1}{H^{2}}-a^{2}r_{0}^{2},\ \ \ \tilde{r}%
_{A-}^{2}=a^{2}r_{0}^{2}.
\end{equation}%
It is evident from equation (37) that the outer trapping horizon will
contract while the inner horizon will expand. It is also interesting to note
that the sum and product of squares of wormhole horizons satisfy $\tilde{r}%
_{A+}^{2}+\tilde{r}_{A-}^{2}=\frac{1}{H^{2}}$, and $\tilde{r}_{A+}^{2}\tilde{%
r}_{A-}^{2}=\frac{a^{2}r_{0}^{2}}{H^{2}}$, respectively. From Eq. (36) the
trapping horizon $\tilde{r}_{A+}$ and $\tilde{r}_{A-}$ coincide at the
extreme case leading to the extreme trapping horizon $\tilde{r}_{A}=1/\sqrt{2%
}H$. Moreover in this case, the wormhole parameters leading to $\dot{a}^{2}=%
\frac{1}{4r_{0}^{2}},$ which upon integration gives $a(t)=\pm \frac{t}{2r_{0}%
}.$ Here the constant of integration is taken to be zero. It shows that the
wormhole is expanding uniformly if $a(t)>0$ and contracting if $a(t)<0$.
Also the naked wormhole is obtained if the discriminant $%
1-4H^{2}a^{2}r_{0}^{2}<0$, which yields $\dot{a}>\frac{1}{2r_{0}}.$

\section*{(2+1)-Wormhole Thermodynamics}

In this section, we discuss the horizon thermodynamics of (2+1) evolving
wormhole. Hayward first introduced a formalism for defining thermal
properties of black holes in terms of measurable quantities. This formalism
also works for the dynamical black holes which consistently recover the
results obtained by global considerations using the event horizon as in the
static case. A fascinating and rather surprising feature emerges if one
recognize that the static wormhole (where it is not possible to infer any
property similar to those found in black holes physics using global
considerations) reveals thermodynamic properties analogous to the black
holes if one considers the local quantities$.$ It is also important to note
that the non-vanishing surface gravity at the wormhole throat characterized
by a non-zero temperature for which one would expect that wormhole should
emit some sort of thermal radiation. Since in our case, the wormhole is
defined to be traversable therefore any matter or radiation could pass
through it from one spacetime to an other (or from a region to another of
the same spacetime) which finally come out in the latter spacetime. One can
discriminate the phenomenon in which a radiation travel through the wormhole
which follows a path allowed by classical general relativity whereas ,
thermal radiation from the horizon is essentially quantum mechanical
process. Therefore, in case of no matter or radiation would pass through the
wormhole throat classically, the existence of a trapping horizon of the
evolving wormhole would produce a quantum thermal radiation. We assume that
the entropy associated with the outer trapping horizon $\tilde{r}_{A+},$ is
proportional to the horizon area analogous to the black hole entropies. So
in this case the entropy of the wormhole becomes%
\begin{equation}
S=4\pi \tilde{r}_{A+}.
\end{equation}%
The surface gravity is defined as \cite{Cai}%
\begin{equation}
\kappa =\frac{1}{2\sqrt{-h}}\partial _{a}(\sqrt{-h}h^{ab}\partial _{b}\tilde{%
r}),
\end{equation}%
where $h$ is the determinant of metric $h_{ab}$ (13). The direct calculation
of the surface gravity from Eq. (39) at the wormhole horizon $\tilde{r}_{A+}$
yields%
\begin{eqnarray}
\kappa &=&-\frac{\tilde{r}_{A+}}{2}\Big(\dot{H}+2H^{2}-\frac{a^{2}r_{0}^{2}}{%
\tilde{r}_{A+}^{4}}\Big),  \nonumber \\
&=&-\frac{1}{\tilde{r}_{A+}}\Big(1-\frac{\dot{\tilde{r}}_{A+}}{2H\tilde{r}%
_{A+}}\Big)\Big(1-\frac{2a^{2}r_{0}^{2}}{\tilde{r}_{A+}^{2}}\Big).
\end{eqnarray}%
The factor $-\frac{1}{\tilde{r}_{A+}}\Big(1-\frac{\dot{\tilde{r}}_{A+}}{2H%
\tilde{r}_{A+}}\Big)$, in (41) is the general expression for the surface
gravity of FRW universe while the second factor $\Big(1-\frac{2a^{2}r_{0}^{2}%
}{\tilde{r}_{A+}^{2}}\Big)$ appears due to the wormhole geometry. It is to
be noted that when $r_{0}$ approaches to zero, the expression for the
surface gravity reduces to the expression for FRW universe. It is important
to mention that the surface gravity vanishes at extreme case which is
consistent with the extreme black hole case.

Now we can compute the horizon temperature $T$ for the evolving wormhole
with the help of the relation%
\begin{equation}
T=-k/2\pi .
\end{equation}

It is evident from Eq. (41) that the horizon temperature may negative,
however, one can avoid any possible physical implication of it by claiming
that it would be a problem only if one is at the trapping horizon. In
addition, one must consider that the radiation infalling in one of the
wormhole mouths will travel classical trajectory to re-appear at the other
mouth of the wormhole as an outgoing radiation. Similarly, the same process
would also\ take place at the other end. So finally we see an outgoing
radiation which surely be unavoidable with negative temperature in the
universe whenever a wormhole is investigated. Moreover, as mentioned earlier
that the phantom energy would be a source to construct the traversable
wormhole and it has been argued that the phantom energy may be characterized
by a negative temperature \cite{PFG}. Therefore, the above result (41)
indicates that wormholes sustained by phantom energy should emit thermal
radiations with negative temperature analogous to the black holes
constructed by the ordinary matter with positive temperature.

After substituting the value of surface gravity (40) in Eq. (41), we obtain

\begin{equation}
T=\frac{1}{2\pi \tilde{r}_{A+}}\Big(1-\frac{\dot{\tilde{r}}_{A+}}{2H\tilde{r}%
_{A+}}\Big)\Big(1-\frac{2a^{2}r_{0}^{2}}{\tilde{r}_{A+}^{2}}\Big).
\end{equation}%
We have determined the expression for the temperature of the evolving
wormhole under consideration. Now our next step is to rewrite the
Friedman-like equation (33) as a first law of thermodynamics. So for this
let us first we write down the Friedman-like equation (33) at the trapping
horizon $\tilde{r}_{A+}.$ To do so, we put the value of $\dot{H}$ from Eq.
(35) in (33), to get%
\begin{equation}
\Big(1-\frac{2a^{2}r_{0}^{2}}{\tilde{r}_{A+}^{2}}\Big)d\tilde{r}_{A+}=\pi H%
\tilde{r}_{A+}^{3}(\rho +p)dt.
\end{equation}%
On multiplying both sides of the above equation by a factor $(1-\frac{\dot{%
\tilde{r}}_{A+}}{2H\tilde{r}_{A+}})$ and arranging the terms, we get
\begin{equation}
\frac{1}{2\pi \tilde{r}_{A+}}\Big(1-\frac{\dot{\tilde{r}}_{A+}}{2H\tilde{r}%
_{A+}}\Big)\Big(1-\frac{2a^{2}r_{0}^{2}}{\tilde{r}_{A+}^{2}}\Big)d(4\pi
\tilde{r}_{A+})=2\pi H\tilde{r}_{A+}^{2}\Big(1-\frac{\dot{\tilde{r}}_{A+}}{2H%
\tilde{r}_{A+}}\Big)(\rho +p)dt.
\end{equation}%
From equation (44), one can recognize that the term on the left hand side is
$TdS$, where $S=4\pi \tilde{r}_{A+}$, is the entropy of the wormhole. So the
above equation reduces to
\begin{equation}
TdS=2\pi H\tilde{r}_{A+}^{2}\Big(1-\frac{\dot{\tilde{r}}_{A+}}{2H\tilde{r}%
_{A+}}\Big)(\rho +p)dt.
\end{equation}%
Now we consider the total matter-energy $E=\rho A$, surrounded by the
trapping horizon $\tilde{r}_{A+}$ of the evolving wormhole. Taking the
differential of $E$ and using the energy conservation relation, we get
\begin{equation}
dE=2\pi \tilde{r}_{A+}\rho d\tilde{r}_{A+}-2\pi \tilde{r}_{A+}^{2}H(\rho
+p)dt.
\end{equation}%
Using Eqs. (45) and (46), we finally achieve
\begin{equation}
dE=-TdS+WdA,
\end{equation}%
where $W=(\rho -p)/2$ is the work density which is defined by $W=-\frac{1}{2}%
T^{ab}h_{ab}$ \cite{SA1,SA2,SA3,SA4,Cai1}. The work term $WdA$ can be
interpreted as the work done against the pressure at the trapping horizon.
The expression (47) is recognized as the unified first law of thermodynamics
\cite{SA1,SA4,Cai2}. The negative sign appears in the first term of right
hand side of Eq. (47) is mainly appears due to the exotic matter which gets
energy from the spacetime itself. Hence one can interpret that the change in
the internal energy $dE$ of the evolving wormhole equals the sum of the
energy removed from the system (wormhole) plus the energy consumed in
performing work done against the pressure. In short, by employing the
entropy proportional to the horizon area together with the total matter
energy density within the trapping horizon of the wormhole, we are able to
show that the Friedman-like equation (33) can be expressed as a
thermodynamic identity. So the notions of temperature and entropy can be
associated with the trapping horizon of the wormhole analogous to the
apparent horizon of FRW universe.

\section*{Conclusion}

It has been shown that the field equations in various theory of gravities
can be recast as a first law of thermodynamics at the horizons of a class of
spacetime geometries. An enormous work dealing with the horizon
thermodynamics is available \cite{Cai,Cai2}. On the other, Hayward has shown
that these geometries (black holes) can be investigated for the
thermodynamic characteristics by taking the trapping horizons into account.
In the Hayward formalism, the black holes and wormhole spacetimes can be
treated on equal footing in order to study their dynamical properties on the
basis of trapping horizons. So our aim has been to visualize the
thermodynamic properties of the evolving wormholes at the trapping horizons.
In this regard, it is shown that the field equations of (2+1)-dimensional
evolving wormhole can be expressed as a first law of thermodynamics $%
dE=-TdS+WdA,$ at the trapping horizon. Here $E=\rho A$ is the total energy
of the matter inside the horizon. Here $W=(\rho -p)/2$ and $A=\pi \tilde{r}%
_{A}^{2}$ are the work density and area respectively. We also discussed the
embedded diagram to study the evolutionary behavior of the evolving wormhole.%
\newline
~\newline \textbf{Acknowledgement:}

M. Akbar is grateful to the organizing committee of the 3rd Algerian
Workshop on Astronomy and Astrophysics where this work was
presented. M. Akbar also expresses his gratitude to the organizing
committee for travel and financial support.

\end{document}